 \documentclass[preprint2]{aastex}

\usepackage{graphicx}
\usepackage{ulem}
\usepackage{color}

\def\ergscm{erg~s$^{-1}$~cm$^{-2}$}
\def\ergs{erg~s$^{-1}$}
\def\2rxp{2RXP~J130159.6-635806}
\def\nuband{$3-78$~keV}
\def\nustar{\textit{NuSTAR}}
\def\psrb{PSR\,B1259-63}

\slugcomment{Not to appear in Nonlearned J., 45.}

\shorttitle{2RXP short title}
\shortauthors{Krivonos et al.}

\begin{document}


\title{NuSTAR discovery of an unusually steady long-term spin-up of 
the Be binary 2RXP J130159.6-635806}


\author{
Roman~A.~Krivonos\altaffilmark{1,2},
Sergey~S.~Tsygankov\altaffilmark{3,2},
Alexander~A.~Lutovinov\altaffilmark{2},
John~A.~Tomsick\altaffilmark{1},
Deepto~Chakrabarty\altaffilmark{4},
Matteo~Bachetti\altaffilmark{5,6},
Steven~E. Boggs\altaffilmark{1}, 
Masha~Chernyakova\altaffilmark{7,8},
Finn~E.~Christensen\altaffilmark{9},
William~W.~Craig\altaffilmark{10,1}, 
Felix~F\"urst\altaffilmark{11},
Charles~J.~Hailey\altaffilmark{12},
Fiona~A.~Harrison\altaffilmark{11},
George~B.~Lansbury\altaffilmark{13},
Farid~Rahoui\altaffilmark{14,15},
Daniel~Stern\altaffilmark{16},
William~W.~Zhang\altaffilmark{17}
}

\altaffiltext{1}{Space Science Lab, University of California, Berkeley, CA 94720;}
\altaffiltext{2}{Space Research Institute, Russian Academy of Sciences, Profsoyuznaya 84/32, 117997 Moscow, Russia;}
\altaffiltext{3}{Tuorla Observatory, Department of Physics and Astronomy, University of Turku, V\"ais\"al\"antie 20, FI-21500 Piikki\"o, Finland;}
\altaffiltext{4}{MIT Kavli Institute for Astrophysics and Space Research, Cambridge, MA 02139, USA}
\altaffiltext{5}{Observatoire Midi-Pyr\'{e}n\'{e}es, Universit\'{e} de Toulouse III  -- Paul Sabatier, 31400 Toulouse, France}
\altaffiltext{6}{CNRS, Institut de Recherche en Astrophysique et Planetologie,  31028 Toulouse, France}
\altaffiltext{7}{Dublin City University, Dublin 9, Ireland}
\altaffiltext{8}{Dublin Institute for advanced studies, 31 Fitzwilliam Place, Dublin 2, Ireland}
\altaffiltext{9}{DTU Space - National Space Institute, Technical University of Denmark, Elektrovej 327, 2800 Lyngby, Denmark}
\altaffiltext{10}{Lawrence Livermore National Laboratory, Livermore, CA 94550}
\altaffiltext{11}{Cahill Center for Astronomy and Astrophysics, California Institute of Technology, Pasadena, CA 91125}
\altaffiltext{12}{Columbia Astrophysics Laboratory, Columbia University, New York, NY 10027}
\altaffiltext{13}{Department of Physics, University of Durham, South Road, Durham DH1 3LE}
\altaffiltext{14}{European Southern Observatory, K. Schwarzschild-Str. 2, 85748 Garching bei M\"unchen, Germany}
\altaffiltext{15}{Department of Astronomy, Harvard University, 60 Garden street, Cambridge, MA 02138, USA}
\altaffiltext{16}{Jet Propulsion Laboratory, California Institute of Technology, Pasadena, CA 91109}
\altaffiltext{17}{NASA Goddard Space Flight Center, Greenbelt, MD 20771}




\begin{abstract}
  We present spectral and timing analysis of \nustar\ observations of
  the accreting X-ray pulsar \2rxp. The source was serendipitously
  observed during a campaign focused on the gamma-ray binary \psrb\
  and was later targeted for a dedicated observation.  The spectrum
  has a typical shape for accreting X-ray pulsars, consisting of a
  simple power law with an exponential cutoff starting at $\sim 7$~keV
  with a folding energy of $E_{\rm fold}\simeq 18$~keV.  There is also
  an indication of the presence of a 6.4~keV iron line in the spectrum
  at the $\sim3\sigma$ significance level. \nustar\ measurements of the
  pulsation period reveal that the pulsar has undergone a strong
  and steady spin-up for the last 20 years. The pulsed fraction is
  estimated to be $\sim80\%$, and is constant with energy up to
  40~keV. The power density spectrum shows a break towards higher
  frequencies relative to the current spin period. This, together with
  steady persistent luminosity, points to a long-term mass accretion
  rate high enough to bring the pulsar out of spin equilibrium.

\end{abstract}



\section{Introduction}

2RXP J130159.6-635806, first discovered by the \textit{ROSAT}
observatory, was later rediscovered in hard X-rays by the
\textit{INTEGRAL}/IBIS telescope and designated with the name IGR\,J13020-6359
\citep{2006ApJ...636..765B,2006AstL...32..145R}.  The first
comprehensive analysis of the temporal and spectral X-ray properties of this
source was done by \cite{masha2005} using data from
the \textit{ASCA}, \textit{BeppoSAX}, \textit{INTEGRAL} and
\textit{XMM-Newton} observatories. In particular, \textit{XMM-Newton} data
showed coherent pulsations with a period of around 700~s. Joint spectral analysis of
\textit{XMM-Newton} and \textit{INTEGRAL} data demonstrated that the
spectral shape is very typical for accretion-powered X-ray pulsars
(namely, an absorbed power law with a high-energy cut-off).  

Based on 2MASS archival data, \cite{masha2005} proposed that the
binary companion to 2RXP J130159.6-635806 is a Be star at a distance
of 4--7~kpc.  This suggestion was later confirmed by
\cite{2013A&A...560A.108C}, who reported the presence of emission
lines of He\,I $\lambda$2.0594 $\mu$m and Br(7--4) $\lambda$2.1663
$\mu$m, which are typical for a Be star. The spectral type of the
optical counterpart was determined to be B0.5Ve.  The orbital period
of the binary remains unknown.

X-ray pulsars in binary systems with Be companions (BeXRPs) typically
manifest themselves as transient sources through either Type I (periodic
flares related to the periastron passage) or Type II outbursts (powerful
rare transient events), or a combination of both
\citep[e.g.,][]{2011Ap&SS.332....1R}. 2RXP\,J130159.6-635806 
shows several differences from a standard transient BeXRP. Specifically, it 
has a relatively low persistent flux, a long pulse period, and it does not 
demonstrate either Type I or Type II outbursts. \cite{masha2005}, however, 
did report some variability of its X-ray flux.

Therefore, there are substantial reasons to consider 2RXP\,J130159.6-635806
as a member of the subclass of persistent BeXRPs \citep{reig1999}. So far,
only a few members of this relatively small category of objects have been
studied in detail: 4U~0352+309/X~Persei, RX~J0146.9+6121/LS~I~+61~235, 
RX~J0440.9+4431, and RX~J1037.5-564 \citep{haberl1998,reig1999}.

In this paper, we present results of a comprehensive analysis of the 
temporal and spectral properties of 2RXP\,J130159.6-635806 in a broad 
energy range, finding some properties that are very unusual for BeXRPs.
All errors are quoted at the 90\% confidence level unless otherwise stated.




\section{Observations}

\begin{figure*}[ht]
\includegraphics[scale=0.93,bb=47 273 555 517,clip]{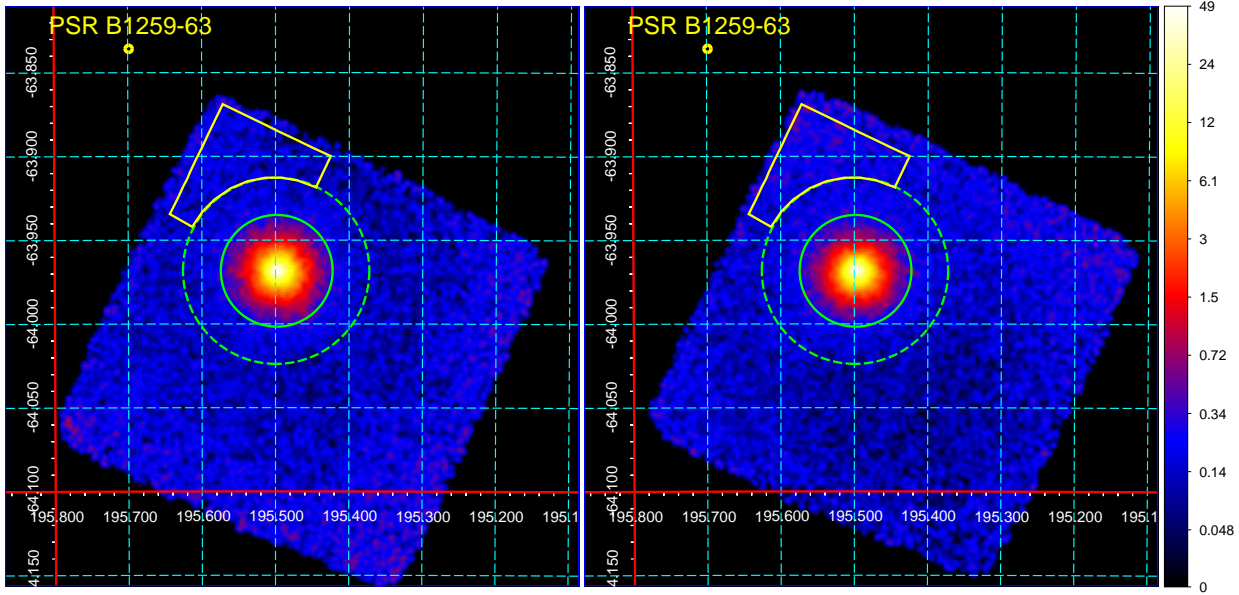}
\caption{Exposure-corrected FPMA (left) and FPMB (right) images of
  \2rxp for the 4th observation in \nuband\ band.
  The images have been smoothed by a Gaussian kernel with 3 pixel
  width ($1 {\rm\ pix}=3''$). Each image is color-coded in logarithmic
  scale. The color bar on the right shows the units of the images
  expressed in $10^{-4}$~cts~pix$^{-1}$~s$^{-1}$. The solid green
  circle ($120''$ radius) and the yellow shape denote regions for the
  source and background extraction, respectively. The dashed green
  circle demonstrates an angular distance of $200''$ from the
  target. The position of the nearby bright source \psrb\ is
  indicated.}
\label{fig:map}
\end{figure*}

\2rxp\ was initially serendipitously observed with \textit{Nuclear
  Spectroscopic Telescope Array (NuSTAR)} \citep{fiona2013} during
observations of the gamma-ray binary \psrb, with three data sets taken
in 2014 May-June (Chernyakova et al., in prep.). In one observation,
\2rxp\ appears in the corner of $\sim 13' \times 13'$ \nustar's field
of view (FOV), and, in two more, the source is at the extreme edge of
the FOV.  Despite the large off-axis angles, the \nustar\ data were
successfully used to extract coherent pulsations. This motivated the
\textit{NuSTAR} team to trigger an on-axis 30~ks observation of \2rxp\
in order to obtain high-quality data for spectral and timing
analysis. Table~\ref{tab:log} lists the \textit{NuSTAR} observations
used in this work.

\begin{table*}
\begin{center}
\caption{{\it NuSTAR} observations\label{tab:log}}
\begin{tabular}{cclccccccccc}
\tableline\tableline
Seq. & Obs. & \multicolumn{1}{c}{Start Time}
& Exp.  &
\multicolumn{2}{c}{Net Count Rate\tablenotemark{a}} & Period\tablenotemark{b} \\
Num.    & ID & \multicolumn{1}{c}{[ UTC ]} & [ ks ]  &
    \multicolumn{2}{c}{ [ counts s$^{-1}$ ] } & [ s ] \\
\tableline
1 & 30002017004  & 2014-05-04 10:01 & 33.3 &
$(1.49\pm0.02)\times10^{-1}$ & $(2.94\pm0.03)\times10^{-1}$ & $643.68 \pm 0.02$\\
2 & 30002017008 & 2014-06-02 19:21 & 26.4 & $(2.00\pm0.12)\times10^{-2}$ &
 $(7.41\pm0.20)\times10^{-2}$ & $643.64 \pm 0.30$ \\
3 & 30002017010 & 2014-06-14 17:21 & 29.1 &
$(7.05\pm0.19)\times10^{-2}$ & $(1.62\pm0.03)\times10^{-1}$ & $643.14 \pm 0.18$ \\
4 & 30001032002 & 2014-06-24 00:06 & 31.6 & $1.342\pm0.007$  &
$1.239\pm0.007$ & $642.90 \pm 0.01$ \\
\tableline
\end{tabular}
\tablenotetext{a}{Net count rate in \nuband\ band for FPMA and FPMB
  extracted from a circular region with a radius of $120''$.}
\tablenotetext{b}{Measured pulsation period for \2rxp.}
\tablecomments{The target for the first three observations was \psrb,
  which placed \2rxp\ offset by $9.55'$ from the optical axis. The 4th
  observation was taken with \2rxp\ on-axis.}
\end{center}
\end{table*}


\textit{NuSTAR} carries two co-aligned identical X-ray telescopes
operating in wide energy band from 3 to 79 keV with angular resolution
of 18'' (FWHM) and half-power diameter (HPD) of 58''. Spectral
resolution of $400$~eV (FWHM) at $10$~keV is provided by independent
focal planes for each telescope, usualy referred as focal plane module
A and B (FPMA and FPMB).

{\it NuSTAR} data can have systematic positional offsets as high as $10''$.  
Prior to extraction, we therefore corrected the
world coordinate system (WCS) of the event files for the four observations
to match the \psrb\ and \2rxp\ centroid positions based on cataloged coordinates.

Since the \nustar\ PSF has wide wings \citep{fiona2013,an2014}, we
investigated how the surface brightness of the source changes with
radius in order to determine regions where the source dominates over
the background. We found that \2rxp\ is well above background within
$120''$ \citep[$\sim92\%$ of PSF enclosed energy; see,
e.g.,][]{an2014}, and that the background count rate can be robustly
measured at radii $\geq$ $200''$ from the source. Taking this into
consideration, we defined the corresponding extraction regions shown
in Fig.~\ref{fig:map} for the 4th (on-axis) observation. The other
three observations, for which \2rxp\ is far off-axis, have been
treated similarly.  Following {\it NuSTAR} recommended standard
practice, we chose the background regions to be on the same detector
chip as the source.

\section{Timing analysis}

\2rxp\ is a known source of coherent X-ray pulsations at a period of
$\sim 700$~s with an average spin-up rate of $\dot\nu \simeq
2\times10^{-13}$~Hz~s$^{-1}$ \citep{masha2005}. We performed timing
analysis of the \textit{NuSTAR} data using the {\sc xronos}
\citep[epoch folding tool efsearch;][]{1983ApJ...266..160L} after
barycentering the data with {\it barycorr}.  For each NuSTAR
observation, the pulse period and its uncertainty were calculated
following the procedure described in
\citep{2013AstL...39..375B}. Namely, a large number ($10^4$) of source
light curves were simulated, the pulse period of each one was
  determined with efsearch, and the distribution of the corresponding
pulse periods was constructed. The mean value of this distribution and
its standard deviation were taken as the pulse period and its
$1\sigma$ uncertainty, correspondingly.  Table~\ref{tab:log} lists
period results derived from the FPMA and FPMB combined light
curves. The inset of Fig.~\ref{fig:period} shows the evolution of the
spin period as a function of time.


As seen from Table~\ref{tab:log} and Fig.~\ref{fig:period}, all four
\nustar\ datasets are suitable for pulsation detection. It is also
quite evident that periods recorded over the time span of 50 days are not
consistent with each other, clearly showing a decrease in the
period.  We utilized the 1st and 4th \nustar\ observations, which have
the most accurate period measurements and also span the full duration
of the {\it NuSTAR} coverage, to measure a period derivative of $\dot
P = -0.0154(5)$~s/day, equivalent to $-1.78(6)\times10^{-7}$
s~s$^{-1}$, or $\dot\nu \simeq 4.3\times10^{-13}$~Hz~s$^{-1}$. This is
in agreement with the \cite{masha2005} spin-up measurement of the
second interval of their data, after the `break' at MJD $\sim51900$
($\dot\nu \simeq 4\times10^{-13}$~Hz~s$^{-1}$). This is quite
remarkable since there is almost a decade between the period
measurements.


\subsection{Pulse period long-term evolution}

\begin{figure}
\vbox{
\includegraphics[width=\columnwidth,bb=20 280 515 675,clip]
{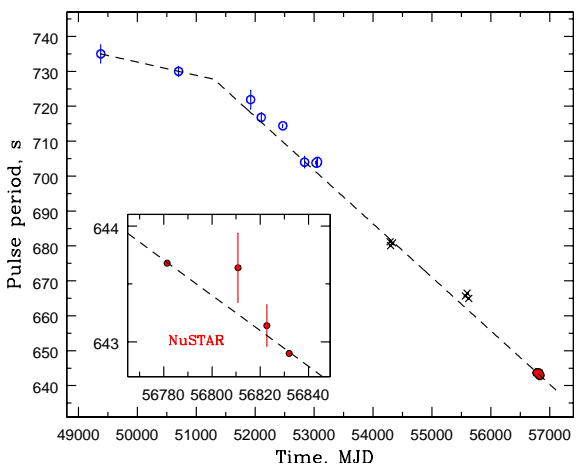}}
\caption{Evolution of the pulse period as a function of time. Blue
  circles show period measurements published by \cite{masha2005} using
  \textit{ASCA}, \textit{BeppoSAX} and \textit{XMM-Newton} data;
  values obtained in this work are shown by black crosses and red
  points for data from \textit{XMM-Newton} and \textit{NuSTAR},
  respectively. Dashed lines represent linear fits to the period
  evolution with two different spin-up rates (see text for the
  details).}\label{fig:period}
\end{figure}

\2rxp\ regularly fell into the FOV of various X-ray telescopes thanks
to extensive observational campaigns dedicated to \psrb\, which is
located only $9.55'$ away. This allows us to investigate the long-term
evolution of the pulse period. We analyzed the
\textit{XMM-Newton} \citep{jansen2001} archival data from 2007-2011
using the procedures described by \cite{masha2005} and Science
Analysis Software (SAS) version 14.0.0.  The list of selected
\textit{XMM-Newton} observations with corresponding period
measurements are shown in Table~\ref{tab:xmm}.

\begin{table}
\begin{center}
\caption{\textit{XMM-Newton} observations\label{tab:xmm}}
\begin{tabular}{cccccccccccc}
\tableline\tableline
 Obs. & Start Time
& Exp.  & Period \\
 ID & [ UTC ] & [ ks ]  & [ s ] \\
\tableline
0504550501  & 2007-07-08 12:01 & 14.1 & $681.20\pm0.15$ \\
0504550601  & 2007-07-16 19:59 & 55.3 & $680.06\pm0.10$ \\
0504550701  & 2007-08-17 08:38 & 11.4 & $680.87\pm0.20$ \\
0653640401  & 2011-01-06 17:37 & 19.9 & $665.65\pm0.20$ \\
0653640501  & 2011-02-02 18:59 & 26.3 & $666.50\pm0.20$ \\
0653640601  & 2011-03-04 05:59 & 12.9 & $664.95\pm0.20$ \\
\tableline
\end{tabular}
\end{center}
\end{table}

Fig.~\ref{fig:period} presents the long-term evolution of the period,
showing that the \2rxp\ neutron star has undergone very strong and
steady spin-up during the last 20 years.  The first available value of
spin period, measured in 1994, is 735~s \citep{masha2005}. The
most recent measurements by \nustar, from 2014, show the period to be
around 643~s (Table~\ref{tab:log}). This means that during the last
$\sim20$ years the spin period decreased by $\sim92$ s, corresponding
to a mean spin-up rate of
$\sim1.4\times10^{-7}$~s~s$^{-1}$. Fig.~\ref{fig:period} shows a
change in the average spin-up rate, first reported by
\cite{masha2005}.

We approximated the long-term period evolution with a linear function
with one change in slope.  A fit shows that the break occurred at MJD
$51300\pm217$ (mid 1999) with a spin-up rate before and after the
break of $(4.3\pm2.7)\times10^{-8}$ s s$^{-1}$ and
$(1.774\pm0.003)\times10^{-7}$ s s$^{-1}$, respectively. As seen from
the fit parameters, the spin-up rate becomes significantly higher
after the break. As noted above, the inset of Fig.~\ref{fig:period}
shows that the \nustar\ data points fit the long-term spin-up rate
with remarkably high precision.

Similar behavior was observed for the X-ray pulsar GX~1+4, which
showed steady spin up for more than a decade \citep[see,
e.g.,][]{1997ApJS..113..367B,2012A&A...537A..66G}.  However, GX~1+4
belongs to the subclass of accreting X-ray pulsars known as symbiotic
X-ray binaries (SyXBs). For BeXRPs, persistent sources
typically demonstrate pulse periods that are relatively stable.
Examples include the population of Be systems in the Small Magellanic
Cloud \citep{klus2014} and the well-known low luminosity Galactic
system X~Persei \citep{lut2012}.  Transient BeXRPs show strong spin-up
during Type I and Type II outbursts \citep{1997ApJS..113..367B} with
significant spin-down episodes in between \citep[see,
e.g.,][]{postnov2015}.  Therefore, 2RXP J130159.6-635806 is a unique
source among the BeXRPs because it demonstrates steady and high
long-term spin-up with a relatively low and stable luminosity.

\subsection{Pulse profile and pulsed fraction}

\begin{figure}
\vbox{
\includegraphics[width=\columnwidth,bb=90 140 500 710,clip]
{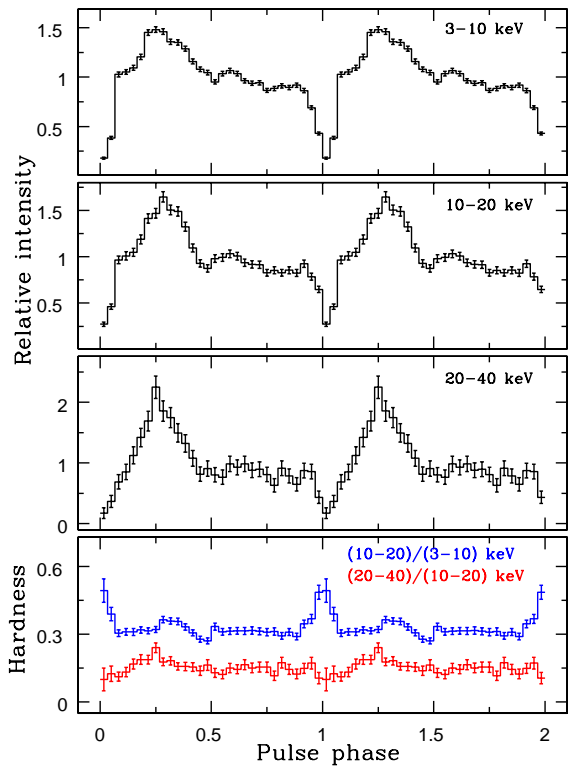}}

\caption{\textit{NuSTAR} \2rxp\ pulse profiles in
  three energy bands (3--10, 10--20 and 20--40 keV), normalized by the
  mean flux. The lower panel shows two hardness ratios, (10--20 keV)/(3--10 keV)
  and (20--40 keV)/(10--20 keV) in blue and red,
  respectively. The profiles are shown twice for clarity.}\label{fig:pprof}
\end{figure}

Pulsar pulse profiles and their evolution with luminosity and
energy band depend on the geometrical and physical properties of the
emitting regions in the vicinity of the neutron star. In
Fig.~\ref{fig:pprof}, the \textit{NuSTAR} pulse profiles of \2rxp\ are shown in three different energy
bands: 3--10, 10--20 and 20--40 keV. The lower panel shows ``soft''
((10--20)/(3--10) keV) and ``hard'' ((20--40)/(10--20) keV) hardness
ratios.

At all energies, the pulse profile can roughly be divided into one 
main peak at phases 0.0--0.5 and two smaller peaks at phases 
0.5--0.75 and 0.75--1.0.  The main feature that changes with energy 
is the depth of the minimum at phase 0.0.  As shown in the lower 
panel of Fig.~\ref{fig:pprof}, while the ``hard'' hardness 
ratio is almost constant, the ``soft'' hardness ratio shows a maximum at phase 
0.0 due to an increase in the depth of the minimum at 3--10 keV. Such 
behavior is caused by differences in the source 
spectrum with pulse phase (see Section 4).

\begin{figure}
\vbox{
\includegraphics[width=\columnwidth,bb=20 275 515 675,clip]
{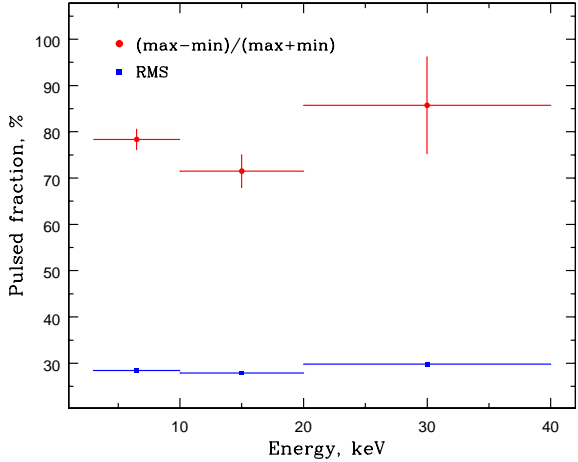}}

\caption{The pulsed fraction (red circles) and relative $RMS$ 
  (blue squares) of 2RXP\,J130159.6-635806 obtained with \textit{NuSTAR}
  as a function of energy.}\label{fig:ppfrac}
\end{figure}

Fig.~\ref{fig:ppfrac} shows the pulsed fraction as a function of
energy.  The pulsed fraction is defined as
$\mathrm{PF}=(I_\mathrm{max}-I_\mathrm{min})/(I_\mathrm{max}+I_\mathrm{min})$,
where $I_\mathrm{max}$ and $I_\mathrm{min}$ are the maximum and
minimum intensities in the pulse profile, respectively. Defined in
this way, the pulsed fraction is very high (about 80\%). The
  alternative way to characterize the pulsed fraction is the relative
  Root Mean Square (RMS), which can be calculated using the following
  equation:
\begin{equation}\label{rms}
RMS=\frac{\Big(\frac{1}{N}\sum_{i=1}^N(P_i-<P>)^2\Big)^{\frac{1}{2}}}{<P>},  
\end{equation}

where $P_i$ is the background-corrected count rate in a given bin of
the pulse profile, $<P>$ is the count rate averaged over the pulse
period, and N is the total number of phase bins in the profile ($N=30$
in our analysis). The $RMS$ deviation obtained in this way reflects
the variability of the source pulse profile in a manner that is not
sensitive to outliers like the narrow features seen in the profile of
\2rxp. Therefore, this quantity has a value of around 30\% that is
much lower than the classically determined pulsed fraction and also
independent of the energy band (see Fig.~\ref{fig:ppfrac}).


It is interesting to note that in contrast to the majority of X-ray
pulsars \citep{lut09}, \2rxp\ does not show an increase in the pulsed
fraction at higher energies. Such uncharacteristic behavior was
previously observed for another persistent BeXRP -- RX\,J0440.9+4431
\citep{2012MNRAS.421.2407T}. On the other hand, Fig.~\ref{fig:pprof}
shows that the pulsed fraction increases somewhat with energy if
$I_\mathrm{min}$ is defined from the the pulse plateau rather than
from the pulse minimum. In other words, the peak-to-plateau difference
slightly grows with energy.

\subsection{Power Spectrum}
\label{section:powerspec}

The observed 20-year strong and steady spin-up of \2rxp\ reveals the
existence of a long-term accelerating torque, which indicates that the
binary interaction must lead to regular accretion over a decade-long
time scale, although this could certainly be episodic (e.g., at
periastron passages). The torque can be transferred by matter
  accreted from either the disc around the neutron star or a stellar
  wind from the optical counterpart.  Unfortunately, there is no
  strong observational evidence allowing us to distinguish between
  these two different accretion channels.  In both scenarios, this
  process is defined mainly by the mass accretion rate and the
magnetic field strength \citep[see, e.g.,][]{1979ApJ...234..296G}.


\begin{figure}
\includegraphics[width=\columnwidth,bb=55 275 550 675,clip]{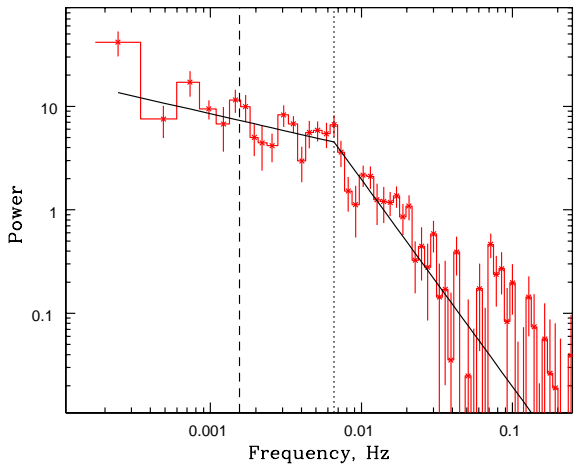}
\caption{The noise power spectrum of \2rxp\ obtained with the
  \textit{NuSTAR} data in observation 30001032002 (red
  histogram; the 4th observation). The solid line shows a broken power law model with the fitted
  value of the break frequency, 0.0066 Hz (vertical dotted line).  The power law slope
  above the break is fixed to --2.  The position of the pulse frequency 
  (0.0015 Hz) is shown by the vertical dashed line.}\label{fig:pds}
\end{figure}


Due to the unknown distance of \2rxp, the luminosity and mass
accretion rate are highly uncertain; the magnetic field is also
unknown since no cyclotron line is found in the energy spectrum
(Sect.~\ref{section:spec}).  However, some qualitative conclusions
about the interaction between the accretion disk and the neutron star
magnetosphere can be made from the noise power spectrum of the X-ray
pulsar.

According to the ``perturbation propagation'' model, stochastic
variations of viscous stresses in the accretion disc cause variations
of the mass accretion rate
\citep{1997MNRAS.292..679L,2001MNRAS.321..759C}.  This, in turn,
results in a specific shape of the Power Density Spectrum (PDS) of the
emerging light curve. Namely, it will appear as a power law with slope
--1 to --1.5 (but the exact value is not well established for X-ray
pulsars) up to the break frequency, which is the highest frequency
that can be generated in the accretion disk
\citep{1997MNRAS.292..679L}.

In the case of a highly magnetized neutron star, this maximal
frequency is limited by the Keplerian frequency at the magnetospheric
radius, above which one can expect a cutoff in the source PDS
\citep{2009A&A...507.1211R}. If the source stays in spin equilibrium
(corotation), the cutoff frequency will coincide with the spin
frequency of the pulsar, whereas, in the case of spin-up (increased
mass accretion rate), the magnetosphere will be squeezed and
additional noise will be generated at higher frequencies. If the mass
accretion rate is known (i.e., the distance to the source is known),
this property of the PDS can be used to estimate the magnetic field
strength of the neutron star
\citep{2009A&A...507.1211R,2012MNRAS.421.2407T,2014A&A...561A..96D}.
The appearance of the break in the PDS does not necessarily indicate
that accretion is from a disk.  There are wind-accreting sources in
spin equilibrium also showing a break in their PDSs around the pulse
frequency \citep{1993ApJ...411L..79H}. However, the evolution of the
PDS shape aw a function of mass accretion rate in such systems is not
well studied.

In Fig.~\ref{fig:pds}, we show the PDS of \2rxp\ obtained with the
\textit{NuSTAR} data in the 4th observation after subtracting the
pulse profile folded with the measured period from the light
curve. The solid line represents the fitting model in the form of a
broken power law. The measured break frequency is 0.0066~Hz (shown by
dotted line), and it is clear that it is shifted towards higher
frequencies relative to the spin frequency in this observation
(0.0015~Hz; shown by dashed vertical line).  The power-law slope above
the break frequency is fixed at --2 \citep{2009A&A...507.1211R}. The
best-fit value of the slope below the break is $-0.33$. Given the steady persistent luminosity, we conclude that
the spin-up observed during last $\sim$$20$ years is caused by a
long-term mass accretion rate that is high enough to squeeze the
magnetosphere inside the corotational radius. This finding confirms
the uniquness of \2rxp\ among the other X-ray pulsars in binary
systems with Be companions.

\section{Spectral analysis}
\label{section:spec}

We used {\it nuproducts}, a part of the NuSTARDAS package, to extract
source and background spectra and to generate \nustar\ response matrix
(RMF) and effective area (ARF) files for a point source.  In our
analysis we utilized the most recent calibration database (CALDB),
version 20150316. The extracted spectra were then grouped to have more
than 20 counts per bin using the {\it grppha} tool from the HEAsoft
6.15.1 package. We fit the \nustar\ spectra using {\sc xspec} version
12.8.1 \citep{xspec}.

\subsection{Pulse phase-averaged spectroscopy}
\label{section:spe}

According to measurements from the \textit{ASCA} and
\textit{XMM-Newton} observatories, the spectrum of \2rxp\ is
characterized by a moderate absorption value $N_{\rm
  H}=(2.48\pm0.07)\times10^{22}$ cm$^{-2}$, which is stable over a
dozen years \citep{masha2005}. This value was obtained by 
approximating the source spectra as a power law model modified by
interstellar absorption ({\sc wabs} model in the {\sc XSPEC} package).
Note that it is just slightly higher than the value of the
interstellar hydrogen absorption $(1.7-1.9)\times 10^{22}$ cm$^{-2}$
determined by \cite{dickey1990} in the direction of \2rxp.

\begin{figure}
\vbox{
\includegraphics[width=\columnwidth,bb=48 225 548 690,clip]
{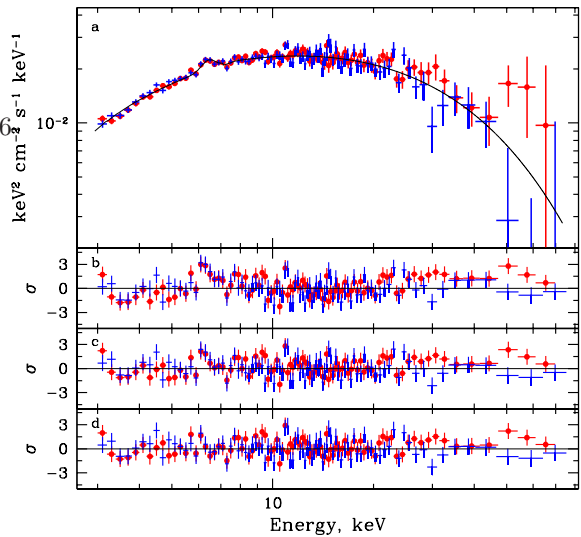}}

\caption{ {\it Top panel (a):} Pulse phase-averaged spectrum of \2rxp\
  obtained with \textit{NuSTAR} during the 4th observation. Red points
  and blue crosses show module A and B data, respectively. The black
  line represents the best fit by the `highecut' model with an iron
  emission line at 6.4~keV. {\it Bottom panels (b,c,d)} show
  corresponding residuals from the models `cutoff', `highecut', and
  `highecut' with the 6.4~keV line, respectively (see text and
  Table~\ref{tab:spectr}).}\label{fig:spectr_onaxis}
\end{figure}

In the first three \nustar\ observations \2rxp\ serendipitously
appeared highly offset from the optical axis, and at the large
off-axis angles the absolute flux measurements can be affected by
inaccurate PSF positioning and subsequent inapropriate weighting in
the spectrum extraction procedure.  This leads to large systematics for
the extracted spectra.  Therefore, we restrict detailed analysis of
the source spectrum to the data obtained in the 4th (on-axis)
observation, which provides high-quality data.

In general, the spectrum of \2rxp\ has a shape which is typical for
accreting pulsars in binary systems, showing a hard spectrum with an
exponential cutoff at high energies. Fig.~\ref{fig:spectr_onaxis} (a)
presents the phase-averaged spectrum approximated with the most
suitable spectral model determined below. We initially modeled data
with a cutoff power-law model ({\sc cutoffpl} in the {\sc XSPEC}
package)
\begin{equation}
AE^{-\Gamma}e^{-\frac{E}{E_{\rm fold}}},
\end{equation}
where $\Gamma$ is the photon index, $E_{\rm fold}$ is the
characteristic energy of the cutoff (e.g., the folding energy), and
$A$ is a normalization.  This model was modified by interstellar
absorption in the form of the {\sc wabs} model. As the working energy
range of the {\it NuSTAR} observatory begins at 3 keV, it is not very
sensitive to measuring low absorption columns.  Therefore, in the
following analysis, the interstellar absorption was fixed to the value
$N_{\rm H}=2.48\times10^{22}$ cm$^{-2}$ measured by
\citet{masha2005}. Note that we simultaneously fitted spectra obtained
by both {\it NuSTAR} modules.  To take into account the uncertainty in
their relative calibrations, which may be even more of a concern for the
observations where the source is highly offset from the optical axis,
we added a cross-calibration constant between the modules.  The
fitting parameters for different data sets are shown in
Table~\ref{tab:spectr}.

The `cutoff' model approximates the source spectrum relatively well
with $\chi^{2}=992.18$ for 921 degrees-of-freedom (\textit{dof}; see
Table~\ref{tab:spectr}). Nevertheless, a wave-like structure is
clearly seen in the residual panel Fig.~\ref{fig:spectr_onaxis} (b).
To improve the quality of the fit, we applied another continuum model
in the form of a power-law multiplied by a high-energy cutoff ({\sc
  highecut} model in the {\sc XSPEC} package; \citealt{white83}).
This model can be written as:
\begin{equation}
AE^{-\Gamma}\times \left\{ \begin{array}{ll}
                       1 & \mbox{($E \leq E_{\rm cut}$)}\\
                       e^{-(E-E_{\rm cut})/E_{\rm fold}} & \mbox{($E > E_{\rm cut}$),} \end{array} \right.
\end{equation}
where $E_{\rm cut}$ is the energy where the cutoff starts. As in the
previous case, we fixed the interstellar absorption value.  Residuals
of modeling the source spectrum with `highecut' are presented in
Fig.~\ref{fig:spectr_onaxis} (c), and best fit parameters are listed
in Table~\ref{tab:spectr}. The `highecut' model significantly improves
the quality of the fit ($\chi^{2}=929.72$ for 920 \textit{dof}). The
high-energy cutoff value $E_{\rm cut}$ is found to be
$6.48_{-0.15}^{+0.22}$~keV, which is significantly lower than the
value of $\sim 25$~keV reported by \cite{masha2005} using simultaneous
\textit{XMM-Newton} and \textit{INTEGRAL} data. The apparent
discrepancy is probably due to the lower statistical quality of the
\textit{INTEGRAL} data and the gap between the energy bands covered by
the \textit{XMM-Newton} and \textit{INTEGRAL} observatories.  An
additional possible problem is that the cutoff energy $E_{\rm
  cut}=6.48$ keV is very close to the energy of the iron fluorescent
line at 6.4 keV.  The simplistic `highecut' model might hide the
presence of the emission line in the source spectrum.  Observed
deviations of the measured spectrum from the model near this energy
argue in favor of this possibility.

To investigate this issue, we added a Gaussian emission line to the
`highecut' model, fixing its energy to 6.4 keV and width to $0.1$~keV,
allowing its normalization to be a free parameter.  This resulted in
an additional improvement of the fit to $\chi^{2}=916.97$ for 919
\textit{dof} for a normalization of $(2.31\pm0.70)\times10^{-5}$ ph
cm$^{-2}$ s$^{-1}$.  The corresponding equivalent width of the line is
$EW=44_{-32}^{+48}$ eV ($3\sigma$ error). We determined the
significance of the line using the {\sc XSPEC} script {\it simftest}
with $4\times10^4$ trials, and found that presumption against the null
hypothesis, or no line required by the data is $3\times10^{-3}$, which
corresponds to $\sim3\sigma$ line detection, assuming a normal
distribution. The residuals of the `highecut' model with the 6.4~keV iron
line are shown in Fig.~\ref{fig:spectr_onaxis} (d). The model itself,
together with spectral data points, is shown in
Fig.~\ref{fig:spectr_onaxis} (a).

\begin{table*}
\begin{center}
\caption{Parameters for the \2rxp\ phase-averaged spectral analysis
  based on \nustar\ observations.\label{tab:spectr}}
\begin{tabular}{ccllccccccccc}
\tableline\tableline
\multicolumn{1}{c}{Seq.} & 
\multicolumn{1}{c}{Model\tablenotemark{a}} & 
\multicolumn{1}{c}{Const\tablenotemark{b}} &
\multicolumn{1}{c}{Photon} &
\multicolumn{1}{c}{$E_{\rm cut}$} & 
\multicolumn{1}{c}{$E_{\rm fold}$}  &
\multicolumn{2}{c}{Flux$^{\rm c}_{2-10{\rm\ keV}}$}  & 
$\chi_{\nu}^2$ (dof) \\
Num.& &  &\multicolumn{1}{c}{index} & \multicolumn{1}{c}{[ keV ]}  & \multicolumn{1}{c}{[ keV ]} 
& FPMA & FPMB & &  &  \\
\tableline
1 & HI & $1.24\pm0.02$ &  $1.40_{-0.14}^{+0.07}$ & $7.47_{-1.55}^{+0.86}$ & $17.43_{-1.29}^{+1.79}$  & $2.16_{-0.38}^{+0.12}$ & $2.68_{-0.48}^{+0.13}$ & 1.06 (458) \\
2 & HI & $0.61\pm0.04$ &  $1.37$ (frozen) & $5.72_{-1.90}^{+0.65}$ & $15.49_{-1.73}^{+2.62}$  & $3.29_{-1.47}^{+0.26}$ & $2.01_{-0.95}^{+0.04}$ & 0.92 (124) \\
3 & HI & $0.92\pm0.03$ &  $1.37$ (frozen) & $5.68_{-0.62}^{+0.53}$ & $17.48_{-1.35}^{+1.45}$  & $2.11_{-0.13}^{+0.08}$ & $1.95_{-0.10}^{+0.06}$ & 0.81 (264) \\
4 & CU & $1.00\pm0.01$ &  $1.04\pm0.03$ &  & $12.62\pm0.48$ & $3.79_{-0.10}^{+0.05}$  & $3.79_{-0.11}^{+0.05}$ & 1.07 (921) \\
  & HI & $1.00\pm0.01$ &  $1.32\pm0.03$ & $6.48_{-0.15}^{+0.22}$ & $16.78_{-0.56}^{+0.68}$ & $3.79_{-0.10}^{+0.05}$ & $3.79_{-0.11}^{+0.05}$ & 1.01 (920) \\[0.5mm]
 & HI+Fe\tablenotemark{d} & $1.00\pm0.01$ &  $1.37\pm0.04$ & $6.94_{-0.42}^{+0.50}$ & $17.96\pm0.87$ &  $3.79_{-0.10}^{+0.05}$ & $3.79_{-0.11}^{+0.05}$ & 1.00 (919) \\[0.5mm]

\tableline
\end{tabular}
\tablenotetext{a}{The {\sc xspec} spectral model used. `CU': {\sc
    wabs*cutoffpl}, `HI': {\sc wabs*powerlaw*highecut}, `HI+Fe': {\sc
    wabs(powerlaw*highecut+gau)}.}  \tablenotetext{b}{Constant factor
  of the FPMB spectrum relative to FPMA.}  \tablenotetext{c}{The
  absorbed flux in units of $10^{-11}$\ergscm.}  \tablenotetext{d}{The
  corresponding 6.4 keV iron line parameters are: $\sigma=0.1$~keV
  (fixed), normalization $(2.31\pm0.70)\times10^{-5}$ ph cm$^{-2}$
  s$^{-1}$, and line equivalent width $EW=44_{-32}^{+48}$ eV
  ($3\sigma$ error).}
\end{center}
\end{table*}


Considering the moderate spectral resolution of the
\nustar\ observatory and the low significance of the detected iron
line, we investigated the possibility of the presence of an iron
fluorescence line in the \textit{XMM-Newton} spectrum of \2rxp. As
mentioned above, the X-ray pulsar \2rxp\ was observed with
\textit{XMM-Newton} many times during programs studying \psrb.
For our purposes, we chose the two observations with the longest
exposures -- ObsID 0092820301 ($\sim$41.2 ksec) and ObsID 0504550601
($\sim$55.3 ksec). As for the \nustar\ observations, we modeled the
spectra of \2rxp\ including the Gaussian line at 6.4 keV.  For both of
the \textit{XMM-Newton} observations, we did not find a significant
improvement in the fits when the iron line was added and obtained a
conservative upper limit ($3\sigma$) for the equivalent width of the
iron line of 110 eV, which is consistent with the \nustar\ results.

Despite the fact that the first three \nustar\ observations were made
at large offset angles, and the corresponding statistics are
significantly lower than for the 4th observation, some useful spectral
information can still be extracted. For these observations we used the
same `highecut' model. Due to low statistics and poor fit we fixed
power-law slope in the 2nd and 3rd data set at $\Gamma=1.37$ value
measured in the 4th observation. As seen from Table~\ref{tab:spectr},
where the best fit parameters are listed, the principal parameters --
power-law slope ($\Gamma$), cut-off energy ($E_{\rm cut}$), and
folding energy ($E_{\rm fold}$) -- of the first three high-offset
observations are in general agreement with 4th (on-axis) observation
modeled with `highecut'. 

The estimated $2-10$~keV flux of \2rxp\ is about
$3\times10^{-11}$~\ergscm\ during our observations, which is in
agreement with the value of $(2-3)\times10^{-11}$~\ergscm\ measured by
\cite{masha2005}. An absence of strong transient activity from \2rxp\
on long time scales is confirmed with the RXTE/ASM and Swift/BAT
instruments. Note that \cite{masha2005} reported on a flaring episode,
but the flux of the flare was relatively low raising by a factor of a
few (up to $\sim10^{-10}$~\ergscm).

Finally, in order to check for the possible presence of a cyclotron
absorption line in the source spectrum, we modified the best-fit model
{\sc (wabs*(powerlaw*highecut+gau))} by including an absorption feature in the
form of a Lorentzian optical depth profile \citep[{\sc cyclabs} model
in {\sc XSPEC};][]{1990Natur.346..250M}. The search procedure was
performed following the prescription from
\cite{2005AstL...31...88T}. Namely, we varied the energy of the line
over the range between 5 and 50 keV with 3-keV steps and the line
width between 2 and 12 keV with 2-keV steps, leaving the line depth as
a free parameter.  We did not find strong evidence for a cyclotron
line in the spectrum of \2rxp (no trials gave a significance higher
than $\sim2\sigma$).

\subsection{Pulse phase-resolved spectroscopy}
\label{section:phase}

\begin{figure}
\vbox{
\includegraphics[width=\columnwidth, bb=55 220 548 691,clip]
{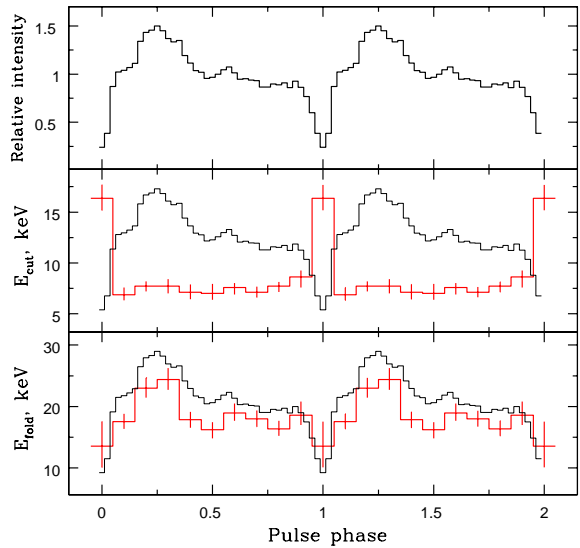}}

\caption{Parameters of the best fit model {\sc (wabs*(powerlaw*highecut+gau))}
  as a function of pulse phase in the 4th (on-axis) observation. {\it Top:}
  the black histogram shows the pulse profile in the entire energy
  range (duplicated in the other two panels).  {\it Middle and bottom,
    respectively:} cutoff energy ($E_{\rm cut}$) and folding energy
  ($E_{\rm fold}$).}\label{fig:spectr_resol}
\end{figure}

In order to study the evolution of the source spectrum over the pulse
period, we performed pulse phase-resolved spectroscopy using the data
from the 4th observation.  The period was divided into 10 phase bins
with zero phase coinciding with the main minimum in the pulse profile.
According to the pulse phase-averaged spectral analysis, the fitting
model was chosen in the form of {\sc
  (wabs*(powerlaw*highecut+gau))}. However, due to much lower
statistics, the photon index was fixed at the value from the averaged
spectrum ($\Gamma=1.37$). We selected this parameter due to its
virtual constancy over the pulse in our preliminary analysis of the
same data. The $N_{\rm H}$ value is not very well constrained by the
\nustar\ data; however we checked that it is consistent with being
constant in phase, and we fixed it as well.

The results are shown in Fig.~\ref{fig:spectr_resol}. The average
pulse profile of \2rxp\ across the entire \textit{NuSTAR} energy range
is presented in the upper panel.  The lower panels demonstrate the
behavior of the two free spectral parameters: the cut-off energy
($E_{\rm cut}$) and the folding energy ($E_{\rm fold}$). The cut-off
energy is quite stable over the pulse except at the pulse minimum
where its value increases approximately by a factor of two.  The
folding energy demonstrates an apparent correlation with the pulse
intensity, which is probably caused by increasing a spectral hardness
around the pulse maximum. Such behaviour of the spectral parameters
over the pulse can explain the corresponding behavior of the hardness
ratios constructed from the pulse profiles in different energy bands,
previously shown in Fig.~\ref{fig:pprof}. The $(20-40)/(10-20)$~keV
hardness ratio demonstrates correlation with the pulse intensity and
the $(10-20)/(3-10)$~keV ratio peaks at the pulse minimum right in
place where $E_{\rm cut}$ shifts from $\sim7$ to
$\sim16$~keV. Finally, it is necessary to note that the observed
behavior of spectral parameters with the pulse phase can be caused by
both physical and artificial reasons (in particular, due to
limitations of available data, the adopted spectral model, and which
model parameters were fixed). More data are required to confidently
constrain all the spectral parameters and trace their behavior with
the pulse phase.

\section{Summary}

We summarize the results of spectral and timing analysis of
serendipitous and dedicated \nustar\ observations of the accreting 
X-ray pulsar \2rxp\ in 2014 May-June.

\begin{itemize}
\item The source demonstrates strong pulsations with a period of
  $\sim640$~s. The $\sim80\%$ pulsed fraction is measured to be
  constant with energy up to 40~keV.

\item The pulse profile is virtually independent of energy and can
  roughly be divided into one main peak at phases 0.0--0.5 and two smaller
  peaks at phases 0.5--0.75 and 0.75--1.0. The only feature that is
  changing with energy is the depth of the main minimum at phase 0.

\item The measured period shows a significant change over the $\sim50$~day
  time span of the \nustar\ observations, which is consistent with a
  spin-up rate of $\dot\nu \simeq 4.3\times10^{-13}$~Hz~s$^{-1}$. This 
  rate is in remarkable agreement with measurements made by
  \cite{masha2005} almost a decade ago.

\item Together with the results of \cite{masha2005}, the \textit{XMM-Newton} 
  data taken in 2007 and 2011, and the \nustar\ observations, we show a
  long-term spin-up trend of the source during the last 20 years. During the 
  last 15 years, the source has undergone a strong and steady spin-up rate at 
  the level of $\dot P = (1.774\pm0.003)\times10^{-7}$ s s$^{-1}$. 

\item The power density spectrum of the source shows a clear break at
  0.0066~Hz, which is higher than the period frequency of
  0.0015~Hz. This fact, together with the steady persistent luminosity
  of the source, implies that the spin-up observed during the last
  20~years is likely caused by a long-term mass accretion rate high enough
  to squeeze the magnetosphere inside the corotational radius, which
  makes \2rxp\ unique among the other X-ray pulsars in binary systems
  with Be companions.

\item The phase-averaged spectrum of the source has a typical shape
  for accreting neutron stars in binary systems, in particular, for
  X-ray pulsars, and demonstrates an exponential cutoff at high
  energies. Our best-fit model contains an absorbed power-law with
  $\Gamma\simeq1.4$ modified by a high-energy spectral drop with a
  cut-off energy of $E_{\rm cut}\simeq7$~keV and a folding energy of
  $E_{\rm fold}\simeq18$~keV. The spectrum also shows $\sim3\sigma$
  evidence for an iron 6.4~keV emission line, with an improvement in
  the fit when it is included.

\item The observed flux corresponds to an {\it unabsorbed} luminosity
  in the range $\sim(8-26)\times10^{34}$~\ergs, assuming a source
  distance between 4 and 7 kpc \citep{masha2005}.  These luminosity
  values imply that the source is a member of the subclass of
  persistent low luminosity Be systems \citep{2011Ap&SS.332....1R}.

\item The phase-resolved spectroscopy shows some differences in source
  spectrum with phase. The cut-off energy is very stable over the
  pulse except at zero phase where its value increases by a factor of
  two. The apparent correlation of the folding energy with pulse
    intensity is attributed to the change in hardness of the source
    spectrum with orbital phase.
\end{itemize}


\acknowledgments

This research has made use of data obtained with \nustar, a project
led by Caltech, funded by NASA and managed by NASA/JPL, and has
utilized the NUSTARDAS software package, jointly developed by the ASDC
(Italy) and Caltech (USA). This research has also made use of data
obtained with XMM-Newton, an ESA science mission with instruments and
contributions directly funded by ESA Member States. AL and ST
acknowledge support from Russian Science Foundation (grant
14-12-01287).


{\it Facilities:} \facility{NuSTAR}, \facility{XMM-Newton}.


\begin{thebibliography}{}

\bibitem[\protect\citeauthoryear{An et al.}{2014}]{an2014} An, H., Madsen, K.~K., Westergaard, N.~J., et al.\ 2014, \procspie, 9144, 91441Q
\bibitem[\protect\citeauthoryear{Arnaud}{1996}]{xspec} Arnaud, K.~A.\ 1996, Astronomical Data Analysis Software and Systems V, 101, 17
\bibitem[\protect\citeauthoryear{Bildsten et al.}{1997}]{1997ApJS..113..367B} Bildsten, L.,  Chakrabarty, D., Chiu, J., et al.\ 1997, \apjs, 113, 367 
\bibitem[\protect\citeauthoryear{Bird et al.}{2006}]{2006ApJ...636..765B} Bird, A.~J., Barlow, E.~J., Bassani, L., et al.\ 2006, \apj, 636, 765
\bibitem[\protect\citeauthoryear{Boldin, Tsygankov, \& Lutovinov}{2013}]{2013AstL...39..375B} Boldin P.~A., Tsygankov S.~S., Lutovinov A.~A., 2013, AstL, 39, 375
\bibitem[\protect\citeauthoryear{Chernyakova et al.}{2005}]{masha2005} Chernyakova, M.,  Lutovinov, A., Rodr{\'{\i}}guez, J., \& Revnivtsev, M.\ 2005,  \mnras, 364, 455
\bibitem[\protect\citeauthoryear{Churazov et al.}{2001}]{2001MNRAS.321..759C} Churazov, E., Gilfanov, M., \& Revnivtsev, M.\ 2001, \mnras, 321, 759
\bibitem[\protect\citeauthoryear{Coleiro et al.}{2013}]{2013A&A...560A.108C} Coleiro A., Chaty S., Zurita Heras J.~A., Rahoui F., Tomsick J.~A., 2013, A\&A, 560, AA108
\bibitem[\protect\citeauthoryear{Dickey \& Lockman}{1990}]{dickey1990} Dickey, J.~M., \& Lockman, F.~J.\ 1990, \araa, 28, 215
\bibitem[\protect\citeauthoryear{Doroshenko et al.}{2014}]{2014A&A...561A..96D} Doroshenko, V., Santangelo, A., Doroshenko, R., et al.\ 2014, \aap, 561, AA96
\bibitem[\protect\citeauthoryear{Ghosh \& Lamb}{1979}]{1979ApJ...234..296G} Ghosh, P., \& Lamb, F.~K.\ 1979, \apj, 234, 296
\bibitem[\protect\citeauthoryear{Gonz{\'a}lez-Gal{\'a}n et al.}{2012}]{2012A&A...537A..66G} Gonz{\'a}lez-Gal{\'a}n, A., Kuulkers, E., Kretschmar, P., et al.\ 2012, \aap, 537, AA66 
\bibitem[\protect\citeauthoryear{Haberl et al.}{1998}]{haberl1998} Haberl, F., Angelini, L., Motch, C., \& White, N.~E.\ 1998, \aap, 330, 189 
\bibitem[\protect\citeauthoryear{Harrison et al.}{2013}]{fiona2013} Harrison, F.~A., Craig, W.~W., Christensen, F.~E., et al.\ 2013, \apj, 770, 103
\bibitem[\protect\citeauthoryear{Hoshino \& Takeshima}{1993}]{1993ApJ...411L..79H} Hoshino, M., \& Takeshima, T.\ 1993, \apjl, 411, L79 
\bibitem[\protect\citeauthoryear{Jansen et al.}{2001}]{jansen2001} Jansen, F., Lumb, D., Altieri, B., et al.\ 2001, \aap, 365, L1 
\bibitem[\protect\citeauthoryear{Klus et al.}{2014}]{klus2014} Klus, H., Ho, W.~C.~G.,  Coe, M.~J., Corbet, R.~H.~D., \& Townsend, L.~J.\ 2014, \mnras, 437, 3863 
\bibitem[\protect\citeauthoryear{Krivonos et al.}{2014}]{arches} Krivonos, R.~A., Tomsick, J.~A., Bauer, F.~E., et al.\ 2014, \apj, 781, 107
\bibitem[\protect\citeauthoryear{Leahy et  al.}{1983}]{1983ApJ...266..160L} Leahy D.~A., Darbro W., Elsner R.~F., Weisskopf M.~C., Kahn S., Sutherland P.~G., Grindlay J.~E., 1983, ApJ, 266, 160
\bibitem[\protect\citeauthoryear{Lutovinov \& Tsygankov}{2009}]{lut09}  Lutovinov A.A., Tsygankov S.S., 2009, Astron. Lett., 35, 433
\bibitem[\protect\citeauthoryear{Lutovinov et al.}{2012}]{lut2012} Lutovinov, A., Tsygankov, S., \& Chernyakova, M.\ 2012, \mnras, 423, 1978 
\bibitem[\protect\citeauthoryear{Lyubarskii}{1997}]{1997MNRAS.292..679L} Lyubarskii, Y.~E.\ 1997, \mnras, 292, 679
\bibitem[\protect\citeauthoryear{Masetti et al.}{2006}]{masetti2006} Masetti, N., Pretorius, M.~L., Palazzi, E., et al.\ 2006, \aap, 449, 1139
\bibitem[\protect\citeauthoryear{Mihara et al.}{1990}]{1990Natur.346..250M} Mihara T., Makishima K., Ohashi  T., Sakao T., Tashiro M., 1990, Natur, 346, 250
\bibitem[\protect\citeauthoryear{Postnov et al.}{2015}]{postnov2015} Postnov, K.~A., Mironov, A.~I., Lutovinov, A.~A., et al.\ 2015, \mnras, 446, 1013 
\bibitem[\protect\citeauthoryear{Reig \& Roche}{1999}]{reig1999} Reig P., Roche P., 1999, MNRAS, 306, 100
\bibitem[\protect\citeauthoryear{Reig}{2011}]{2011Ap&SS.332....1R} Reig, P.\ 2011, \apss, 332, 1
\bibitem[\protect\citeauthoryear{Revnivtsev et al.}{2006}]{2006AstL...32..145R} Revnivtsev,  M.~G., Sazonov, S.~Y., Molkov, S.~V., et al.\ 2006, Astronomy Letters, 32, 145
\bibitem[\protect\citeauthoryear{Revnivtsev et al.}{2009}]{2009A&A...507.1211R} Revnivtsev, M., Churazov, E., Postnov, K., \& Tsygankov, S.\ 2009, \aap, 507, 1211
\bibitem[\protect\citeauthoryear{Tsygankov \& Lutovinov}{2005}]{2005AstL...31...88T} Tsygankov S.~S., Lutovinov A.~A., 2005, AstL, 31, 88
\bibitem[\protect\citeauthoryear{Tsygankov, Krivonos, \& Lutovinov}{2012}]{2012MNRAS.421.2407T} Tsygankov S.~S., Krivonos R.~A., Lutovinov A.~A., 2012, MNRAS, 421, 2407
\bibitem[\protect\citeauthoryear{Verner et al.}{1996}]{verner1996} Verner, D.~A., Ferland, G.~J., Korista, K.~T., \& Yakovlev, D.~G.\ 1996, \apj, 465, 487
\bibitem[\protect\citeauthoryear{White et al.}{1983}]{white83} White N., Swank J., Holt S., 1983, \apj, 270, 771
\bibitem[\protect\citeauthoryear{Wilms et al.}{2000}]{wilms2000} Wilms, J., Allen, A., \& McCray, R.\ 2000, \apj, 542, 914




\end{thebibliography}
\end{document}